\documentclass[sigconf]{acmart}

\usepackage{subcaption}
\usepackage{tabularx}
\usepackage{amsmath}
\usepackage{stfloats}
\usepackage[font=bf, skip=6pt]{caption}

\usepackage{multirow}

\usepackage{calc}
\usepackage{ragged2e}
\usepackage{array}

\usepackage{makecell}

\colorlet{lightgray}{gray!15}
\colorlet{darkgray}{gray!45}

\usepackage{todonotes}
\newcommand{\ignore}[1]{}

\AtBeginDocument{%
  }

\copyrightyear{2025}
\acmYear{2025}
\setcopyright{rightsretained}
\acmConference[Conference 2025]{Conference 2025}{2025}{Location}
\acmBooktitle{Conference 2025}

\begin{document}

\title{Exploring Student-AI Interactions in Vibe Coding}

 \author{Francis Geng}
 \orcid{0009-0004-5188-4755}
 \affiliation{%
   \institution{University of California San Diego}
   \city{La Jolla}
   \state{California}
   \country{USA}}
 \email{fgeng@ucsd.edu}

\author{Anshul Shah}
\orcid{}
\affiliation{%
  \institution{University of California San Diego}
  \city{La Jolla}
  \state{California}
  \country{USA}}
\email{ayshah@ucsd.edu}

\author{Haolin Li}
\orcid{}
\affiliation{%
  \institution{University of California San Diego}
  \city{La Jolla}
  \state{California}
  \country{USA}}
\email{hal180@ucsd.edu}

\author{Nawab Mulla}
\orcid{}
\affiliation{%
  \institution{University of California San Diego}
  \city{La Jolla}
  \state{California}
  \country{USA}}
\email{nmulla@ucsd.edu}

\author{Steven Swanson}
\affiliation{%
 \institution{University of California San Diego}
 \city{La Jolla}
 \state{California}
 \country{USA}}
\email{sjswanson@ucsd.edu}

\author{Gerald Soosai Raj}
\affiliation{%
 \institution{University of California San Diego}
 \city{La Jolla}
 \state{California}
 \country{USA}}
\email{asoosairaj@ucsd.edu}

\author{Daniel Zingaro}
\orcid{0000-0002-1568-4826}
\affiliation{%
 \institution{University of Toronto Mississauga}
 \city{Toronto}
 \state{Ontario}
 \country{Canada}}
 \email{daniel.zingaro@utoronto.ca}

\author{Leo Porter}
 \orcid{0000-0003-1435-8401}
 \affiliation{%
 \institution{University of California San Diego}
 \city{La Jolla}
 \state{California}
 \country{USA}}
\email{leporter@ucsd.edu}

\renewcommand{\shortauthors}{Geng et al.}

\begin{abstract}

\textbf{Background and Context.}  
Chat-based and inline-coding-based GenAI has already had substantial impact on the CS Education community.  The recent introduction of ``vibe coding'' may further transform how students program, as it introduces a new way for students to create software projects with minimal oversight.

\noindent \textbf{Objectives.} The purpose of this study is to understand how students in introductory programming and advanced software engineering classes interact with a vibe coding platform (Replit) when creating software and how the interactions differ by programming background. 

\noindent \textbf{Methods.} Interview participants were asked to think-aloud while building a web application using Replit. Thematic analysis was then used to analyze the video recordings with an emphasis on the interactions between the student and Replit.

\noindent \textbf{Findings.} For both groups, the majority of student interactions with Replit were to test common cases in the prototype or use prompts to debug the prototype. Only rarely did students analyze or manually edit code. Prompts by advanced software engineering students were much more likely to include relevant app feature and codebase contexts than those by introductory programming students. 

\end{abstract}

\ccsdesc[500]{Social and professional topics~Computing education}

\keywords{Large Language Models, Vibe Coding, Novice Programmers, Observation Study}

\maketitle

\section{Introduction}
The advent of GenAI has had considerable impact on computing education. It solves course assignments~\cite{denny2023conversing,denny2024computing}, helps students learn through tutoring~\cite{liffiton2023codehelp,denny2024desirable,molina2024leveraging,kazemitabaar2024codeaid}, alters how we teach~\cite{kazemitabaar2025exploring,tankelevitch2024metacognitive}, and changes the skills we need to teach~\cite{porter2024learn, vadaparty2024cs1,lau2023ban}. 

LLMs, such as Copilot and ChatGPT, have changed the workflow of programming for professionals~\cite{Barke_GroundedCopilot} as well as for students~\cite{porter2024learn,vadaparty2024cs1}. As the CS education research community works to identify the impact of GenAI on student learning (with mixed findings~\cite{kazemitabaar2023studying,Prather_WideningGap,tankelevitch2024metacognitive,lucchetti2024substance}), platforms such as Replit~\cite{replit}, Copilot Agents~\cite{copilot_agents}, and Cursor~\cite{cursor} have the potential to alter the landscape even further. 

The term ``vibe coding'' ~\cite{vibetweet, sarkar25, sapkota25} describes creating software with minimal oversight—users focus on the end product rather than systematically reviewing or testing code. Existing work has largely examined expert or professional programmers, leaving a gap in understanding how students engage in this workflow. For the CS education community, if we can understand how students use AI-powered tools for vibe coding, we will be better equipped to help support students to use them effectively. 
  
We observed how students from an introductory programming course and an advanced software engineering course used Replit to vibe-code web applications. Participants completed a think-aloud task using Replit’s AI Agent and Assistant, and their sessions were qualitatively coded to capture key behaviors including prompting, testing, debugging, and code review.


Our findings suggest two main patterns. First, across both groups, students primarily used Replit to test prototypes and debug through prompts, with limited direct engagement in code. Second, advanced students incorporated more code-level reasoning and richer programming context in their prompts. These differences highlight that experienced students interact with AI tools in more sophisticated ways, underscoring the importance of teaching students at all levels to read code and frame prompts with appropriate context.


 
\vspace{-0.2cm}

\section{Background and Literature Review}

\subsection{Research on Vibe Coding}
\label{sec:vibecoding_lit}
In this section, we explore how vibe coding has been defined, summarize the existing research on vibe coding, and compare vibe coding to first-generation coding with generative AI and agentic coding.
 
\subsubsection{What is Vibe Coding?} 

The term ``vibe coding'' was introduced in February 2025 by Andrej Karpathy, a computer scientist and AI researcher who co-founded OpenAI. The following tweet text is the definition of vibe coding as written by Karpathy~\cite{vibetweet}:

\textit{
``There's a new kind of coding I call `vibe coding', where you fully give in to the vibes, embrace exponentials, and forget that the code even exists. It's possible because the LLMs (e.g., Cursor Composer w Sonnet) are getting too good. Also I just talk to Composer with SuperWhisper so I barely even touch the keyboard. I ask for the dumbest things like `decrease the padding on the sidebar by half' because I'm too lazy to find it. I `Accept All' always, I don't read the diffs anymore. When I get error messages I just copy paste them in with no comment, usually that fixes it. The code grows beyond my usual comprehension, I'd have to really read through it for a while. Sometimes the LLMs can't fix a bug so I just work around it or ask for random changes until it goes away. It's not too bad for throwaway weekend projects, but still quite amusing. I'm building a project or webapp, but it's not really coding --- I just see stuff, say stuff, run stuff, and copy paste stuff, and it mostly works."
}

\noindent{}That is, vibe coding involves asking a GenAI tool for code, but not reading that code. To debug, one can paste error messages into the AI, or ask for random changes until it's fixed. Karpathy notes that this approach is good enough for weekend projects.

To be clear, Karpathy (an expert computer scientist and AI researcher) doesn't need to vibe code. But, because it works for him, he uses it. It's okay if vibe coding only \textit{``usually''} fixes the bug, or only \textit{``mostly''} works, because Karpathy can just fix the code in these cases. That is, it's perhaps unsurprising that experts, with deep knowledge of the underlying code, can ride the vibes until something goes wrong, fix the problem, and continue vibing. This leaves us with two questions: \textit{what does vibe coding actually look like when carried out by experts?} And, \textit{what happens when novices and intermediate programmers vibe code?} There is initial research into the first question that we describe next; and answering the second question is a goal of the present study.

\subsubsection{What Do Expert Developers Do When They Vibe Code?}

The most comprehensive study of developers vibe coding is the study by Sarkar and Drosos~\cite{sarkar25}. These authors studied YouTube and Twitch videos of experienced developers who were vibe coding while thinking aloud. The authors chose to analyze those videos where the programmer self-described what they are doing as vibe coding, rather than attempting to impose some criteria of what vibe coding is. That's because the definition of vibe coding is not up to us, but is and will continue to be negotiated by communities of developers who use it in practice.

The authors found that these programmers benefited from their expertise in several ways, such as when evaluating, testing, or manually editing code, and when including detailed technical specifications in their prompts. They are able to rapidly assess code and make immediate judgments about its suitability. The authors use the term ``material disengagement'' to emphasize the increasing distance between developers and their code in a vibe coding workflow. That is, these developers work on their code not directly, but mediated through GenAI.

The authors' qualitative analysis led to nine top-level categories, including what developers want to build with vibe coding, plans for how they'll build it, the vibe coding workflow, prompting strategies, and debugging. As a sample of what the authors found, we highlight just three of their many insights: 1) Some developers begin with expectations of complete success, but often need to temper those expectations to partial (e.g., 80\%) success; 2) Due to its conversational nature, vibe coding often helped these developers go beyond their initial visions. At the same time, vibe coding could be so fast as to lock a developer into a suboptimal plan before they know it; 3) Developers used both single-objective prompts and multi-objective prompts, the latter of which may include directives to the AI about multiple unrelated requirements.

Again, all of these findings are from expert programmers with considerable understanding of programming languages, or the affordances of AI models, or both. In contrast, the novice and intermediate programmers in our study have substantially less experience in software development and AI, making our study a valuable complement to prior work and helping to fill a critical gap in our understanding of vibe coding among non-experts and students.

\subsubsection{Vibe Coding vs. Other AI Programming Workflows}

While definitions of AI-assisted programming workflows are in flux, we do wish to contrast vibe coding against two other workflows in order to further position vibe coding.

First, we distinguish vibe coding from first-generation GenAI programming workflows from 2022 and 2023, where programmers prompt for each function and the AI completes the code~\cite{sarkar25}. At this time, chat interfaces hadn't been integrated into the GenAI tools yet. Vibe coding, as we have described, is much further abstracted from the code, allowing programmers to delegate significantly larger tasks to the AI. That said, it is \textit{not} entirely hands off~\cite{sarkar25}. 

Second, we distinguish vibe coding from ``agentic coding''~\cite{sapkota25}, in which the intention {\it is} to be hands off. The human is not in the loop: \textit{``agentic coding enables autonomous software development through goal-driven agents capable of planning, executing, testing, and iterating tasks with minimal human intervention''}~\cite{sapkota25}. At least, this is how agentic coding is defined for experienced developers. We wonder to what extent our comparably less experienced students' ``vibe coding'' looks like agentic coding.


\subsection{How Programmers Interact with AI Programming Tools}

Though research related specifically to ``vibe coding''  is in its early stages, a larger body of work has studied how programmers interact with generative AI programming tools. In this section, we describe the literature related to how experienced and inexperienced programmers interact with AI tools such as GitHub Copilot and ChatGPT.

\subsubsection{How Experienced Programmers Use AI Tools}

 
Broadly, the software engineering literature on AI programming tools discusses how these tools can change the software development process~\cite{jin2025swe, Chen_LLMChallenges, Ozkaya_Motivation, Requirements_Development}, how developers use AI tools~\cite{shihab2025effects, Barke_GroundedCopilot, Github_survey, Gao_2024}, and challenges they face when using those tools~\cite{Vaithilingam_Usability, Liao_MATCH, Wang_TrustInCopilot, Perry_InsecureCode}. In this section, we focus on how experienced programmers use AI tools. This helps inform how advanced software engineering students (intermediate-level programmers) might interact with AI in a vibe coding platform.

\citeauthor{Barke_GroundedCopilot} conducted an observational study of 20 programmers with a range of prior experience --- from professional to occasional programmers --- to theorize how programmers use GitHub Copilot~\cite{Barke_GroundedCopilot}. Their theory posits that programmers interact with Copilot in two modes: exploration mode, in which the programmer uses Copilot to understand how to get started with a task; and acceleration mode, in which programmers are aware of the necessary steps and use Copilot to speed up implementation~\cite{Barke_GroundedCopilot}. Recent observational studies of how programmers use AI tools support the theory above, shedding light on the prompting and verification behaviors among developers~\cite{Prather_Weird, Liang_UsabilitySurvey, Vaithilingam_Usability}. Notably, \citeauthor{Liang_UsabilitySurvey} found that programmers used various techniques to verify AI-generated code output, such as by scanning the code for keywords, using a compiler to detect issues, executing the code, or examining the code in depth~\cite{Liang_UsabilitySurvey}. In addition, \citeauthor{Vaithilingam_Usability} and \citeauthor{Perry_InsecureCode} both discussed challenges developers face, with \citeauthor{Vaithilingam_Usability} highlighting the difficulty of comprehending and debugging AI-generated code, and \citeauthor{Perry_InsecureCode} finding that programmers who used AI tools were more likely to write insecure code that included more system vulnerabilities~\cite{Perry_InsecureCode}.  Other studies focused on how developers process the output of AI tools. For example, \citeauthor{Liao_MATCH} discussed how users make trust judgments of AI output, engaging in either systematic processing (requiring a careful evaluation of the output) or heuristic processing (involving the use of heuristics to make quick, yet error-prone, trust assessments)~\cite{Liao_MATCH}. 

In the computing education research space, two recent studies by \citeauthor{shihab2025effects}~\cite{shihab2025effects} and \citeauthor{Shah_Copilot}~\cite{Shah_Copilot} studied upper-division CS students' interactions with AI tools while working on existing code bases. \citeauthor{shihab2025effects} conducted a within-subjects experiment and found that students completed the tasks with Copilot 35\% faster, and that Copilot reduced the amount of time programmers spent writing code by 11\%  and the amount of time spent performing web searches by 12\%~\cite{shihab2025effects}. Similarly, \citeauthor{Shah_Copilot} analyzed the prompting strategies of 48 students as they completed a task to add a feature to an open-source code base, showing that students preferred to interact with Copilot chat to comprehend or generate code and reporting a higher trust in Copilot's code comprehension features than its code generation features~\cite{Shah_Copilot}.

The prior studies mentioned in this section describe workflows with AI tools where programmers are still actively interacting with the generated code. For example, \citeauthor{Barke_GroundedCopilot}, \citeauthor{Liang_UsabilitySurvey}, and \citeauthor{Vaithilingam_Usability} highlight how programmers still spent time reviewing and understanding the code output, especially when trying to debug. However, as discussed in Section~\ref{sec:vibecoding_lit}, vibe coding platforms introduce an even greater degree of abstraction between programmers and their code compared to earlier AI tools like Copilot and ChatGPT. This increased separation makes it especially important to study how the more abstract workflow affects intermediate programmers --- such as our advanced software engineering students --- who have enough experience to understand and occasionally review code, but may not be proficient or confident in debugging complex implementations by hand.


\subsubsection{How Novice Programmers Use AI Tools}


Plenty of work has analyzed how introductory students program with AI tools~\cite{lucchetti2024substance, Prather_Weird, Prather_WideningGap, Amoozadeh_CS1, denny2023conversing}, how these tools impact the help-seeking landscape~\cite{Hou_ErodingInteractions, kazemitabaar2024codeaid,  molina2024leveraging, denny2024desirable}, and how to teach in the age of generative AI~\cite{vadaparty2024cs1, porter2024learn, kazemitabaar2024codeaid, lau2023ban, denny2024computing}. Given our study's focus on how novice and intermediate programmers approach vibe coding, we will discuss prior works that have observed how novices interact with various AI tools for programming.

Novice programmers tend to exhibit ineffective prompting and verification strategies, with multiple studies showing that students try to write prompts that would solve the entire problem at once~\cite{Amoozadeh_CS1, Kazemitabaar_CS1}. Studies have also highlighted novices' ineffective verification strategies when working with AI generated code~\cite{Prather_WideningGap, Prather_Weird, Kazemitabaar_CS1}. \citeauthor{Kazemitabaar_CS1} showed that CS1 students only ran the AI generated code 60\% of the time, and 13\% submitted AI-generated code without executing the code at all~\cite{Kazemitabaar_CS1}.  With an emphasis on non-technical end users, \citeauthor{JohnnyCantPrompt} discussed their struggles with AI prompting~\cite{JohnnyCantPrompt}. The authors found a lack of effective prompting strategies among end users, including limited examples or relevant details in their prompts~\cite{JohnnyCantPrompt}. The study also discussed users' tendency to over-generalize the AI tool's behavior, especially when verifying the tool's handling of a particular input (e.g., lack of extensive testing before making a judgment). Though the participants in our study have more programming experience, the results presented by \citeauthor{JohnnyCantPrompt} provide an example of how lay users prompt and verify AI systems. 
Our present study aims to understand the prompting and verification behaviors of novice students who are learning to program, since understanding \textit{how} these students interact with vibe coding platforms can help instructors adapt their teaching to prepare students for a future in which vibe coding and other AI-integrated workflows are a reality.

Studies have aimed to explain why novices struggle when programming with AI. \citeauthor{lucchetti2024substance} attributed novices' ineffective prompting to difficulty incorporating the context that is needed for the AI to generate effective code; and that, when students get caught in unproductive prompting cycles, it's often because they continue making surface-level wording changes rather than adding what's missing~\cite{lucchetti2024substance}. \citeauthor{Prather_WideningGap} and \citeauthor{tankelevitch2024metacognitive} argued that some of the challenges are metacognitive \cite{Prather_WideningGap, tankelevitch2024metacognitive}. For example, novice students struggle to verify the correctness of output, leading them down an incorrect path that may only compound their existing struggles~\cite{Prather_WideningGap}.

In short, the studies above focused on how novices use and struggle with first-generation AI tools that involve more programmer-code interactions than vibe coding. In response, our study takes the first steps in understanding how novice programmers prompt and verify the output of a vibe coding platform that, by design, lowers the interactions between programmers and code.

\vspace{-0.2cm}

\section{Methods}
\subsection{Research Questions}
The research questions for our study are as follows:
\begin{description}
    \item[\textbf{RQ1}] How do students interact with AI tools when engaging in a vibe coding workflow?
    \item[\textbf{RQ2}] How does prior programming experience influence the way students develop software in a vibe coding workflow?
\end{description}

\subsection{Study Context}

\subsubsection{Recruitment}
\label{sec: recruitment}

We recruited students from two computer science courses at a large research-intensive North American university, under an approved Human Subjects protocol. The first course (CS1) primarily enrolls first- and second-year students with some prior programming experience. This 10-week Java course covers variables, conditionals, arrays, recursion, and basic object-oriented programming. The second course, an advanced Software Engineering (SWE) class, emphasizes understanding, managing, and contributing to legacy codebases. It primarily serves final-year students who have completed the core software engineering course, where they learn AGILE development and build mobile or web applications from scratch in a team setting.

\subsubsection{Replit}

\begin{figure}[t]
  \centering
  \includegraphics[width=\columnwidth]{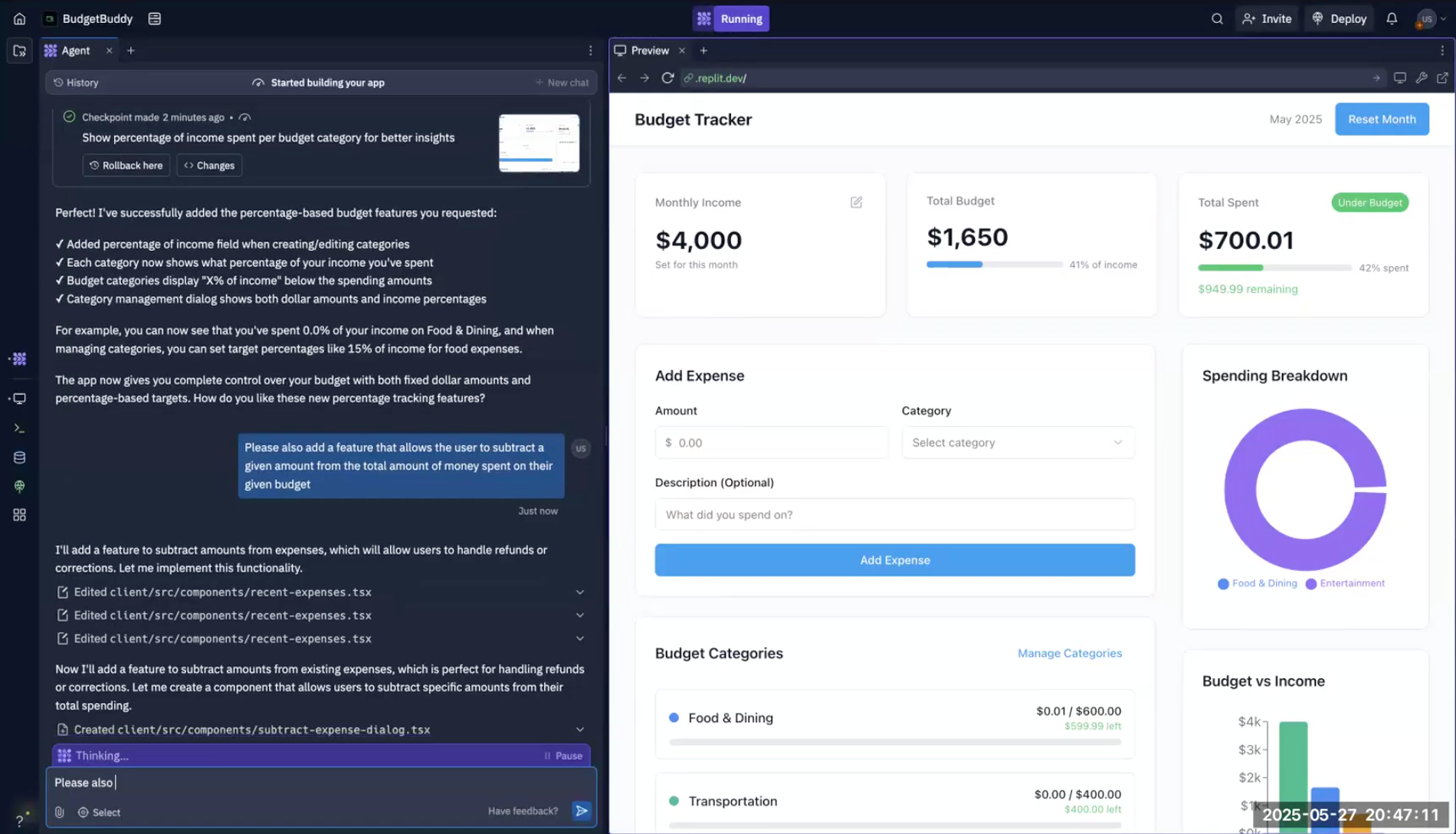}
  \caption{\textbf{Screenshot of the Replit user interface during a task.} }
  \label{fig:rep}
  \vspace{-0.5em} 
\end{figure}

Replit is a browser-based, AI-integrated IDE designed to support a vibe coding workflow~\cite{replit}. It invites users to \textit{``Prompt your app ideas to life — no coding required''}\footnote{https://replit.com/usecases/ai-app-builder}. The interface combines an AI chat panel, where users interact with Replit’s chatbots, and a live preview of the application interface (Figure~\ref{fig:rep}). Initially static, the preview becomes interactive as it reflects the evolving application (\textit{“iterative prototype”}). Unlike traditional IDEs such as VS Code or Eclipse, Replit hides code by default, centering development on AI interaction and prototype feedback rather than manual code editing.

As of May 2025 (when the study was conducted), Replit’s ``AI App Builder'' scaffolded full-stack web applications primarily in TypeScript, unless users requested another framework. We note that TypeScript was unfamiliar to many of our CS1 participants. The platform automatically handled environment setup, file organization, dependencies, and deployment, allowing students to run and interact with a live application within minutes of their first prompt. 

Within this interface, Replit offers two distinct chatbot tools, ``Replit Agent'' and ``Replit Assistant''. The Agent is optimized for building applications from the ground up; it can autonomously scaffold full-stack projects, configure environments, and directly create or modify files based on high-level prompts. In contrast, the Assistant is tailored for working within an existing project. It supports code-level refinement tasks such as debugging, explaining code behavior, refactoring, and minor feature edits. It functions similarly to an in-line coding collaborator: users can ask questions or request specific changes, and the Assistant proposes contextual edits that must be explicitly approved before being applied.

\subsection{Participants}

We recruited 9 students from the CS1 course and 10 students from the SWE course, as defined in Section \ref{sec: recruitment}.

Within the CS1 cohort (CS1\_1–9), 4 students were majoring in computing or related fields and 5 were not. 6 identified as men and 3 as women; 7 identified as Asian or Asian American and 2 as White. 6 students reported confidence in building small class projects, while 3 had only basic programming skills. None had software engineering internship experience involving legacy codebases.

All 10 SWE participants (SWE\_1–10) were computing majors, with 8 men and 2 women; 7 identified as Asian or Asian American, 2 as Chicanx or Latinx, and 1 as White. 4 reported being comfortable with small projects, 4 with intermediate projects, and 2 with large projects or legacy codebases. 6 had prior internships, and 9 aspired to become professional software engineers.

\begin{table*}[h!]
\centering
\caption{Labeling scheme based on student-AI interactions, divided into four major categories.}
\label{tab:labeling_scheme}
\begin{tabularx}{\textwidth}{l X}
\toprule
\textbf{Label} & \textbf{Definition} \\ \hline
\rowcolor{darkgray}
 \textbf{\textit{Interacting with Prototype}} & \\ 
 
 Refresh Prototype & Reloading the tab that displays the prototype \\
  Test Common Case & Interacting with the prototype using typical inputs or actions that reflect normal usage; repeated clicks or filling out a form counts as a single case \\
  Test Edge Case & Interacting with the prototype using unusual, invalid, or boundary inputs such as negative numbers and empty fields \\

\rowcolor{darkgray}
  \textbf{\textit{Writing a Prompt}} & \\
  Debug & Prompting to address an error, bug, or system failure \\
  Add/Remove/Update Core Feature & Prompting to request any changes to a core feature (e.g. adding, checking, deleting, or editing budgets, budget categories, or expenses) that does not have an error or bug\\  
  Add/Remove/Update Non-core Feature & Prompting to request any changes to a non-core feature (e.g. UI elements, spending graphs) that does not have an error or bug\\ 
  Ask Clarification Question & Prompting to ask the AI to explain, define, or elaborate on a concept, term, feature, or technical details\\
  Brainstorm Ideas & Prompting for open-ended ideas or approaches without a clearly defined solution\\ 
  Other & Prompting actions not covered by other labels, such as simulating a test case or responding to the AI’s questions \\

\rowcolor{darkgray}
  \textbf{\textit{Managing Replit Workflow}} & \\ 
  Accept Code Change & Accepting a code modification proposed by the Replit Assistant using the built-in apply-suggestion feature \\ 
  Pause AI During Generation & Interrupting the AI's response before completion by clicking the Replit pause or stop button \\ 
  Resume AI Generation After Pause & Resuming a paused AI response by clicking the Replit continue or resume button \\ 
  Revert to Checkpoint & Reverting the code to a previous version by selecting a checkpoint created by Replit  \\
  Load Preview from Checkpoint & Loading a preview of code from a previous checkpoint without reverting to it \\ 
  Approve Plan with X Additional Features & Approving 0 or more optional features from the Replit Agent’s initial project plan \\


\rowcolor{darkgray}
   \textbf{\textit{Engaging with Code/log}} & \\
   Interpret & Spending significant time analyzing code, console logs, or other development-related outputs. \\
  Edit & Modifying code, console logs, or other development-related outputs \\
\bottomrule
\end{tabularx}
\end{table*}

\subsection{Study Procedure}

The study sessions were conducted one-on-one on Zoom by the first author.
To refine task clarity, we piloted the procedure with six students from the same institution—two from the CS1 course and four upper-division undergraduates with backgrounds similar to the SWE participants. The finalized two-hour study comprised three phases:

\emph{Pre-task training} (30 minutes): Students watched a six-minute tutorial demonstrating the use of Replit Agent and Assistant, accessing the codebase, and interacting with the prototype. The researcher then introduced the think-aloud method via a short live demo, after which students practiced the procedure on a sample task using the Replit platform.

\emph{Development session} (60 minutes, timed): Students were asked to create a personal budget management web app in Replit. The task required four core functions: (1) set a monthly budget; (2) break it into categories (e.g., food, rent, entertainment); (3) record expenses; and (4) compare spending to goals. We made the task description open-ended to encourage diverse student interpretations and resulting application designs. Students could use any Replit features (Agent, Assistant, code editor, console, or preview) and external tools such as ChatGPT. 

The timer began when students opened the task description. Sessions followed the think-aloud protocol, with recordings capturing voice and screen actions; cameras were optional. The researcher observed and took notes throughout.

\emph{Post-task interview} (30 minutes): Each participant completed a short semi-structured interview covering background, initial impressions of Replit, challenges, prior development experience, confidence, and reflection. Example questions included:
\begin{itemize}
    \item What was it like trying to get the Replit agent to understand your goals? 
    \item How did you feel when the agent did not give you what you wanted, even after you felt you gave a clear instruction? 
    \item I noticed that you accessed (or did not access) code during the task — could you share your reasoning? 
    \item Did your prior programming experience influence how you dealt with challenges? 
    \item If you were to re-do the entire activity, what would you do differently from your first attempt? 
\end{itemize}

These questions are not exhaustive but are meant to illustrate the kinds of probes used to support our interpretation of observed behaviors and contextualize the student quotes that we report. We did not perform a full thematic analysis of the interviews or think-aloud data for this paper; instead, we selectively drew on them to provide context. A subsequent paper will present deeper qualitative analysis.

\subsection{Data Analysis}

After transcribing all recordings, two authors conducted a qualitative thematic analysis to identify and label student actions during the development session. They first jointly open-coded one CS1 and one SWE session, then discussed and formed an initial set of labels. Each author independently applied these labels to the remaining 17 videos, engaging in periodic discussions to refine the scheme and resolve discrepancies. Finally, they met to reach a negotiated agreement~\cite{GARRISON20061, sarkar25} on all labels. The finalized labeling scheme is shown in Table~\ref{tab:labeling_scheme}.

To address our research questions, we labeled observable screen-recorded actions in the Replit IDE and, when applicable, external AI tools such as ChatGPT. We also collected associated artifacts—prompts, code files, error messages, and other context—to better interpret behaviors. While our focus was on student actions, we referred to think-aloud utterances and interview responses when needed for context.

\begin{table}[h!]
\centering
\caption{\textit{Writing a Prompt --- Debug} Sublabeling Scheme.}

\begin{tabularx}{\columnwidth}{l X}
\toprule
\textbf{Sublabel} & \textbf{Definition} \\
\midrule
Error Message & Including a partial or full system error message when describing a bug or issue \\ \hline
Failing Case & Describing the input, condition, or steps that led to a failure or error, including what preceded the problem \\ \hline
Code-focused & Referencing specific files, code snippets, or implementation details believed to be related to the issue \\ \hline
Other Details & Providing other relevant information not captured above, such as triggering AI-generated debugging prompts (e.g., \textit{``Troubleshoot this issue''} or ``\textit{``Ask Agent to explain code''}) \\ \hline
Low Context & Debugging prompts with none of the above sublabels selected, showing little or no new information or direction when initiating or continuing a debugging effort\\
\bottomrule
\end{tabularx}
\label{tab:debug_sublabel}
\end{table}

We additionally recorded which AI tool students used for \textit{Writing a Prompt} (Replit Agent, Replit Assistant, ChatGPT) and created sublabels for the \textit{Writing a Prompt—Debug} category (Table~\ref{tab:debug_sublabel}), which represented most prompting interactions and the greatest variation in communication styles. Other prompting categories were less frequent and we did not further subdivide them.

\vspace{-0.2cm}

\section{Results}
\label{sec:results}

\subsection{RQ1: Student-AI Interactions in Vibe Coding}
\label{subsec:rq1}

\begin{figure*}[t]
  \centering
  \includegraphics[width=\textwidth]{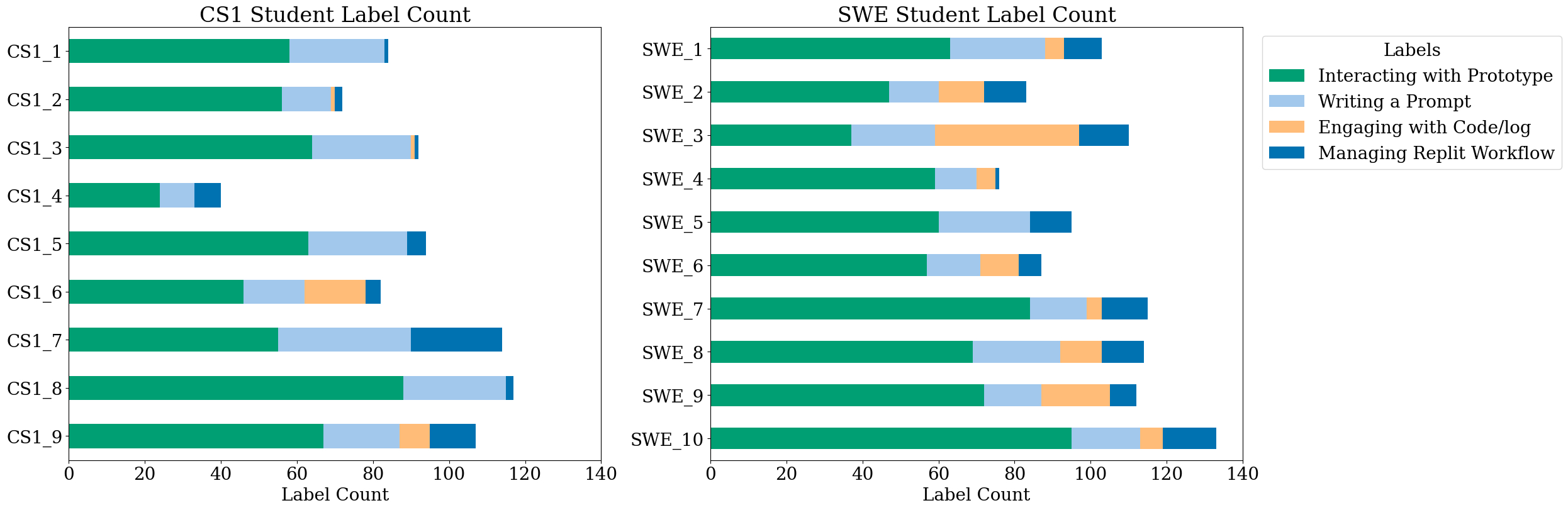}
  \caption{Distribution of labels across all 19 students, grouped by course. Each bar represents a single student, segmented by interaction categories: \textit{Interacting with Prototype}, \textit{Writing a Prompt}, \textit{Engaging with Code/log}, and \textit{Managing Replit Workflow}. SWE\_1 to SWE\_10 are SWE students and CS1\_1 to CS1\_9 are CS1 students.}
  \label{fig:overall_label}
\end{figure*}

\textbf{\textit{Overview.}}
The overall interactions with AI tools across the 19 participants are shown in Figure~\ref{fig:overall_label}. Across all students, the most prevalent label was \textit{Interacting with Prototype} (63.61\% of all labels, n = 1164), followed by \textit{Writing a Prompt} (20.60\%, n = 377), \textit{Managing Replit Workflow} (8.42\%, n = 154), and \textit{Engaging with Code/log} (7.38\%, n = 135). However, differences emerged across cohorts: students in CS1 showed markedly higher proportions of \textit{Writing a Prompt} labels compared to SWE students, whereas \textit{Engaging with Code/log} interactions were more common among SWE students.

\textbf{\textit{Restarters.}}
While prompting was a common behavior across participants, a subset of students (4 out of 19) exhibited a rather dramatic prompting pattern: they restarted the entire project using the Replit Agent mid-task. We term these students \textit{restarters} (SWE\_1, SWE\_2, CS1\_4, CS1\_7). 3 out of the 4 \textit{restarters} restarted primarily to simplify their interaction with the AI, citing overwhelming or ambiguous behavior from earlier prompts. For instance, one student reflected that they \textit{``asked the Replit to do way too many things,''}

 making it difficult to identify specific issues (SWE\_1, interview). Another opted to \textit{``break it down one task at a time''} after experiencing repeated failures (CS1\_7, think aloud). These behaviors suggest that some students use restart strategies not out of failure alone, but as a form of iterative refinement and task decomposition.
 The fourth \textit{restarter} (SWE\_2) retained their original prompt but added a sentence to explicitly request a Flask framework over React due to familiarity and perceived simplicity, and they considered their restart as a means of architectural realignment rather than functional simplification. This finding highlights a critical point: restart decisions reflected metacognitive awareness about the limitations of debugging through prompting and demonstrated students’ adaptive strategies when existing prototypes became unmanageable.

\begin{table}[h!]
\centering
\caption{Frequency of labels for \textit{Interacting with Prototype} actions.}
\begin{tabular}{lrr}
\toprule
\begin{tabular}[c]{@{}c@{}}\textbf{Label}\end{tabular} & \textbf{Count} & \textbf{Percent}\\
\midrule
Test Common Case & 1067 & 91.67\% \\ \hline
Refresh Prototype & 71 & 6.10\% \\ \hline
Test Edge Case  & 26 & 2.23\% \\ \hline
\midrule
\textbf{Total} & 1164 & 100\% \\
\bottomrule
\end{tabular}
\label{tab:interact}
\end{table}
\textbf{\textit{Interacting with Prototype.}}
We examine how students interact with the prototypes generated by the AI tools (Table~\ref{tab:interact}). The overwhelming majority of these interactions (91.67\%) involved testing common use cases, while only 6.10\% involved refreshing the prototype and a mere 2.23\% involved edge case testing. Transcript data suggests that prototype refreshes were typically prompted by technical limitations (e.g., Replit failing to maintain state across pages) rather than as part of a deliberate debugging strategy. Notably, no student wrote or executed unit tests during the study, indicating that their approach to testing was exclusively centered on feature-level behaviors visible through the UI. The absence of structured test practices, combined with minimal edge case coverage, suggests that many students’ testing strategies remained limited in scope and were shaped by immediate usability and tool constraints. These patterns highlight the inherently iterative and sometimes unstable nature of vibe coding, where students often remain occupied with basic interactions and repeated troubleshooting, rather than advancing toward comprehensive feature validation.

\begin{table}[h!]
\centering
\caption{Frequency of labels for \textit{Writing a Prompt} actions.}

\begin{tabular}{lrr}
\toprule
\textbf{Label} & \textbf{Count} & \textbf{Percent} \\
\midrule
Debug & 230 & 61.01\% \\ \hline
Add/Remove/Update Non-core Feature & 63 & 16.71\% \\ \hline
Add/Remove/Update Core Feature & 53 & 14.06\% \\ \hline
Other & 16 & 4.24\% \\ \hline
Ask Clarification Question & 12 & 3.18\% \\ \hline
Brainstorm Ideas & 3 & 0.80\% \\ \hline
\midrule
\textbf{Total} & 377 & 100\% \\
\bottomrule
\end{tabular}
\label{tab:prompt}
\end{table}

\textbf{\textit{Writing a Prompt.}}
To better understand what students' goals are when they are prompting, we analyzed the labels of all prompting behaviors for all 19 students (Table~	\ref{tab:prompt}). The majority of prompts (61.01\%) were used for debugging AI-generated code, followed by modifications to non-core features (16.71\%) and core features (14.06\%). Prompts related to brainstorming, clarification questions, or miscellaneous tasks were relatively rare (less than 5\% each). These data indicate that students primarily engaged with AI tools to troubleshoot and refine partial implementations, rather than to build core functionality
 from scratch. As one student described their strategy: \textit{``...finding [bugs] myself, realizing what the problem was, and then putting it back into the Agent to solve it in like 2 seconds...''} (SWE\_9, interview). This reflects how students often used AI to efficiently resolve implementation issues they had manually identified.

\begin{table}[h!]
\centering
\caption{Frequency of student prompts with different AI tools.}
\begin{tabular}{lrr}
\toprule
\begin{tabular}[c]{@{}c@{}}\textbf{AI Tool}\end{tabular} & \textbf{Count} & \textbf{Percent}\\
\midrule
Replit Assistant & 271 & 50.94\% \\ \hline
Replit Agent & 246 & 46.24\% \\ \hline
ChatGPT & 15 & 2.82\% \\ \hline
\midrule
\textbf{Total} & 532 & 100\% \\
\bottomrule
\end{tabular}
\label{tab:which_ai}
\end{table}

In terms of which AI tools students used for prompting (Table~\ref{tab:which_ai}), the Replit Assistant accounted for a slight majority of prompting interactions (50.94\%), closely followed by Replit Agent (46.24\%). ChatGPT was used only in 2.82\% of the prompt instances, likely reflecting its auxiliary role in the workflow. That is, while multiple AI tools were available, most of the prompting occurred within the embedded Replit interfaces, suggesting that accessibility and immediacy of tooling may play a significant role in shaping student behavior. As one student reflected, \textit{``I didn't really understand the real difference between Agent and Assistant... it kind of felt like they were the same thing''} (SWE\_4, interview), emphasizing how conceptual ambiguity may have contributed to balanced usage. Another student explained, \textit{``I’ll kind of use ChatGPT and Copilot for assistance on making small fixes,''} (SWE\_10, interview) illustrating ChatGPT’s more occasional and supporting role relative to the Replit-native tools.

\begin{table}[h!]
\centering
\caption{Frequency of labels for \textit{Managing Replit Workflow} actions.}
\begin{tabular}{lrr}
\toprule
\begin{tabular}[c]{@{}c@{}}\textbf{Label}\end{tabular} & \textbf{Count} & \textbf{Percent}\\
\midrule
Accept Code Change & 105 & 68.18\% \\ \hline
Approve Plan with X Additional Features & 23 & 14.94\% \\ \hline
Pause AI During Generation  & 12 & 7.79\% \\ \hline
Revert to Checkpoint  & 7 & 4.55\% \\ \hline
Load Preview from Checkpoint  & 5 & 3.25\% \\ \hline
Resume AI Generation After Pause  & 2 & 1.30\% \\ \hline
\midrule
\textbf{Total} & 154 & 100\% \\
\bottomrule
\end{tabular}
\label{tab:workflow_freq}
\end{table}
\textbf{\textit{Managing Replit Workflow.}}
In addition to prompting actions, students interacted with Replit in several other ways that reflect the vibe coding  workflow (Table~\ref{tab:workflow_freq}). The most common action in this category was accepting Replit-proposed code changes (68.18\%), a manual decision-making step mandatory for students working with Replit Assistant. After writing an initial prompt, a student had to approve an AI-generated implementation plan with optional additional features recommended by the Replit Agent (14.94\%) based on the content of that first prompt. Some students also paused (7.79\%) and resumed (1.30\%) the AI generation process, indicating moments of review or reconsideration during code generation. In addition, a few students used Replit’s version control features to either \textit{Revert to Checkpoint} (4.55\%) or \textit{Load Preview from Checkpoint} (3.25\%), actions functionally similar to Git’s \texttt{revert} and \texttt{checkout}, respectively. These behaviors suggest that students occasionally found it necessary to backtrack after a prompt yielded undesirable or destabilizing changes, emphasizing the trial-and-error nature of vibe coding. 

\textbf{\textit{Engaging with Code/log.}}
Among the rare cases where students engaged with code and logs generated by Replit, an overwhelming 90.37\% of the interactions were interpreting and reading code (n = 122), and the remaining actions were direct edits (9.63\%, n = 13). This strong preference for interpretation over modification suggests that students were often hesitant to alter AI-generated code, likely due to limited familiarity with the underlying implementation. The vibe coding workflow may exacerbate this hesitation by distancing students from the logic and structure of AI-generated codebases. As one student explained, \textit{“Because so much of it was just done by the LLM, I had a lesser understanding of the codebase --- rather than what I would do on my own, where I know what each line does”} (SWE\_6, interview).

\subsection{RQ2: Relationship between Programming Experience and Vibe Coding Strategies}\label{subsec:rq2}


All statistical comparisons in this section use two-sample independent t-tests.

\begin{figure}[t]
  \centering
  \includegraphics[width=\columnwidth]{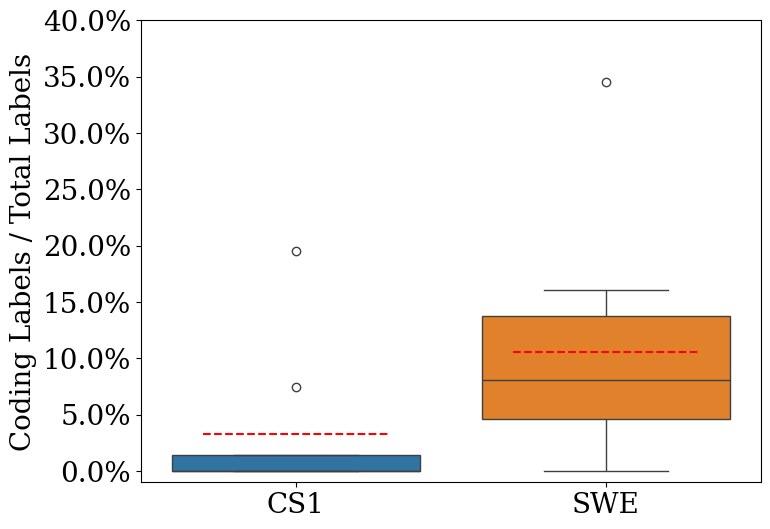}
  \caption{\textbf{Proportion of \textit{Engaging with Code/log} labels per student, normalized by each student’s total behavior labels.} }
  \label{fig:coding}
  \vspace{-0.5em} 
\end{figure}

\textbf{\textit{Engaging with Code/log.}}
To investigate how programming experience influences students' engagement with code, we compared the proportion of \textit{Engaging with Code/log} actions (either interpreting or editting AI-generated code) between students in CS1 and SWE. As shown in Figure~\ref{fig:coding}, SWE students exhibited a higher overall proportion of \textit{Engaging with Code/log} behaviors (M = 10.56\%, SD = 9.83\%) compared to CS1 students (M = 3.27\%, SD = 6.55\%), although the difference approaches but does not reach statistical significance (p = .078).  Among individual behaviors, 9 of 10 SWE students engaged in code interpretation, and 3 of them also edited AI-generated code. By contrast, 4 of 9 CS1 students interpreted code, only 2 of which (CS1\_6 and CS1\_9) edited code. Notably, these two students were the only ones in their cohort who self-identified as capable of developing intermediate-level programs and debugging, suggesting that students with greater programming experience are more likely to engage directly with the structure and content of AI-generated code.

\begin{figure}[t]
  \centering
  \includegraphics[width=\columnwidth]{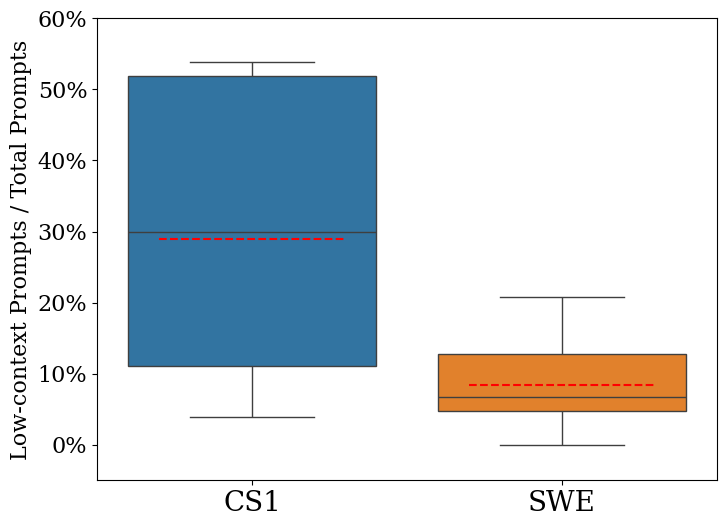}
  \caption{Proportion of \textit{Low Context} prompts per student, normalized by each student’s total number of prompts. }
  \label{fig:low_context}
  \vspace{-0.5em} 
\end{figure}

\textbf{\textit{Writing a Prompt.}}
Students from CS1 and SWE also showed noticeably different behaviors in two aspects when using debugging prompts during vibe coding, characterized as \textit{Low Context} and \textit{Code-focused} (defined in Table~\ref{tab:debug_sublabel}). \textit{Low Context} prompts are often vague and offer limited actionable context, such as \textit{``The buttons appear but they don't work at all,''} or follow-up phrases like \textit{``it’s still not working.''} Figure~\ref{fig:low_context} compares the proportion of low context prompts across the two groups and reveals that CS1 students had a significantly higher proportion of such prompts (M = 28.89\%, SD = 20.66\%) than SWE students (M = 8.39\%, SD = 6.99\%, p \textless .01).

In contrast, \textit{Code-focused} prompts that reference specific segments of AI-generated code or logs were more prevalent among SWE students. For example, one student wrote, \textit{``In line 118 of routes.ts, are you passing the correct data into storage.createStorage?''} (CS1\_6, prompt). Another student wrote, \textit{``In the <Header> component, error occurred 'cannot read properties of undefined (reading `month`)', set the month to be the current month at the time the user is using the web app''} (SWE\_8, prompt). 

These kinds of prompts demonstrate precise understanding and reasoning about the program’s structure and behavior. 7 out of 10 SWE students used code-focused prompts, while only 1 student from the CS1 group (CS1\_6) did so. This difference suggests that students with greater programming experience were more capable of constructing prompts that referenced concrete implementation details, allowing them to more effectively communicate intent and troubleshoot with the AI system.

\textbf{\textit{Interacting with Prototype.}}
We also analyzed the extent to which students conducted edge case testing while \textit{Interacting with Prototype} using uncommon or boundary inputs. 6 of 10 SWE students performed at least one edge case test, compared to 4 of 9 in CS1. While this difference is modest, it may reflect task progression: students who encountered repeated errors in basic functionality often focused on common case testing and did not reach a point where testing edge cases was necessary. Since edge case testing is introduced early in many computing curricula, including this CS1 course, the observed frequency difference between cohorts is less likely due to instruction and more likely tied to whether the student’s implementation had stabilized enough to allow for more nuanced evaluation.

\vspace{-0.2cm}

\section{Discussion}
With the recent arrival of agentic AI and vibe coding, there is considerable discourse around higher-level prompting as a viable practice for creating software. That discourse, however, has been focused mostly on professional software engineers, not students. The fact that vibe coding involves minimal supervision may suggest that someone completely unfamiliar with programming could just dive right into the practice. Our findings raise caution here, in that students from both the introductory and the advanced programming classes relied on their existing programming skills when interacting with AI: prompting to debug, performing feature-level tests, and engaging with code. This suggests that someone with no or very limited programming background may struggle with vibe coding, since even our CS1 students drew on foundational skills to make progress. 

\subsection{Implications for Educators}

In our findings, interacting with the prototype was the most common behavior for both introductory and advanced students. Both groups experimented with the prototype to identify what worked and what didn't, then used that information to ask Replit to correct shortcomings. The fact that students were so quick to move to testing is an encouraging finding for teaching testing in computing courses. However, it is critical to understand the type of testing students engaged in and its implications. Students in our study primarily performed closed box testing---behavioral testing focused on whether features work correctly, without examining or understanding the underlying code. The Replit interface reinforces this approach by intentionally masking code behind one to two button clicks, presenting only the chat prompts and prototype preview by default. While students technically had access to the generated code, most intentionally chose not to engage with it. This stands in contrast to open box testing, such as unit testing, where developers have full access to and understanding of the code structure and can validate that individual functions or components work correctly in isolation. While closed box testing develops valuable skills in 
validating user-facing functionality, it does not build the deeper understanding required for maintaining, debugging, or extending code at scale. Students may graduate competent at verifying that generated code produces correct outputs for specific inputs, yet struggle when asked to integrate their solutions into larger codebases or optimize performance. Educators should explicitly teach students the differences between these testing approaches and when each is appropriate. Furthermore, since Replit and similar tools encourage feature-based testing over unit tests, instructors may need to deliberately scaffold assignments that require students to write unit tests and engage with code structure. 

Our findings also demonstrate that introductory and advanced programming students have different prompting styles. The advanced students are far more apt to communicate with Replit using prompts that contain computational ideas and details than introductory students. These sophisticated prompts suggest, perhaps unsurprisingly, that advanced students drew on more computational ideas when interacting with Replit. While we do not measure prompt success in this study, the difference in style highlights that prompting in vibe coding involves not just phrasing requests, but conveying intent grounded in code structure, logic, and context. This finding supports recent work by Lucchetti et al., who found that programming novices often author prompts lacking important technical details necessary for AI to generate appropriate solutions~\cite{lucchetti2024substance}. The pedagogical implication is clear: {\it prompting effectiveness depends fundamentally on programming knowledge}. Students cannot effectively direct AI to generate appropriate code without understanding what ``appropriate'' means computationally. They may struggle to debug AI-generated errors if they are unaware of the kinds of errors that are possible. They cannot iteratively refine prompts without understanding why initial attempts failed. We argue that computing education should not reduce emphasis on foundational programming concepts in favor of prompting instruction, but rather recognize that strong prompting emerges from strong programming foundations.

\subsection{Limitations}

While our study offers timely insights into how students interact with AI tools in a vibe coding workflow, several limitations must be acknowledged.

\textbf{Sample representation.} 
Our study was conducted at a single North American institution with students recruited from two computing courses, and participation was voluntary. This recruitment strategy may introduce self-selection bias, as students who opted into the study may be more confident, motivated, or interested in AI tools than their peers. Consequently, the behaviors observed may reflect a more engaged or higher-performing subset of the student population. Expanding future studies to multiple institutions and course settings would strengthen the applicability of findings across contexts.

\textbf{Platform specificity.}
Our findings are grounded entirely in student experiences with the Replit platform, which is designed around an AI chat interface and an interactive prototype that deprioritizes code visibility. While this design makes Replit ideal for examining vibe coding, it does not reflect the full spectrum of vibe coding platforms. Other tools (e.g., Cursor, Copilot Agents) may provide greater transparency into generated code, tighter integration with traditional IDE workflows, or different prompting paradigms altogether. As such, our results should not be generalized to \textit{all} vibe coding environments without caution. Future work could incorporate multiple platforms to assess how specific design features shape student behavior.

\textbf{Focus on observed behaviors, not underlying reasoning.} Our analysis centers on students’ observable actions (e.g., prompting, testing, engaging with code) and classifies them using a structured labeling scheme. However, we did not systematically analyze students’ spoken reasoning or reflective interview data to explain \textit{why} those behaviors occurred. While the think-aloud protocol and post-task interviews provide valuable context, we primarily used them to validate labeled behaviors rather than drive deeper thematic or cognitive analysis. As a result, we cannot fully characterize students’ intentions, mental models, metacognitive strategies, or emotional responses. A complementary analysis of verbal data could provide richer insight into students’ goals, frustrations, and strategies while vibe coding.

\textbf{Interaction frequency is not equivalent to time on task.} All quantitative results in this paper are based on counts of interaction labels (e.g., number of prompts, code edits, test cases). However, raw frequency does not capture \textit{how long} students spent on different activities. For instance, a single instance of code interpretation may involve several minutes of analysis, while five quick prompts could span less than a minute in total. As such, interpreting our frequency data as proportional effort or cognitive investment would be misleading. Future work should incorporate duration-based metrics or temporal coding to accurately represent time allocation and sustained engagement with specific activities.

\vspace{-0.2cm}

\section{Conclusion}
This study offers first insights into how computing students interact with a vibe-coding tool (Replit) while creating a functional and useful web app. 
Through qualitatively analyzing observations of how students vibe code, we uncovered students' interaction patterns with Replit and behavioral differences between students in introductory and advanced computer science courses. Our findings include 1) the bulk of student interactions with Replit are closed box testing and prompting to debug, and 2) advanced students are more capable of prompting with computational sophistication and are more likely to interact with the codebase.  Our findings offer insights to educators on the skills that students use when vibe coding while buttressing findings in GenAI broadly --- that testing and debugging remain critical in the AI era~\cite{prather2025beyond, porter2024learn} and that students need training in prompting~\cite{lucchetti2024substance}.
\vspace{-0.2cm}

\section{Acknowledgments}
This work is supported by the National Science Foundation Graduate Research Fellowship Program and NSF Award \#2417374, and by the Google Award for Inclusion Research Program. Any opinions, findings, and conclusions or recommendations expressed in this material are those of the authors(s) and do not necessarily reflect the views of the National Science Foundation or Google. 

\balance
\bibliographystyle{ACM-Reference-Format}
\bibliography{refs}

\end{document}